\documentclass[cameraready]{Interspeech}

\title{Can LLMs Help Localize Fake Words in Partially Fake Speech?}
\author[affiliation={1,2}, orcid=0000-0001-7826-2850]{Lin}{Zhang}
\author[affiliation={1,2}, orcid=0000-0001-8953-7872]{Thomas}{Thebaud}
\author[affiliation={1,2}, orcid=0009-0007-2658-2220]{Zexin}{Cai}
\author[affiliation={1,2}, orcid=0000-0001-5976-0897]{Sanjeev}{Khudanpur}
\author[affiliation={3}, orcid=0000-0002-0611-3634]{Daniel}{Povey} 
\author[affiliation={1,2}, orcid=0000-0002-7449-5726]{Leibny Paola}{Garc\'ia-Perera}
\author[affiliation={1,2}, orcid=0000-0002-5423-7754]{Matthew}{Wiesner}
\author[affiliation={1,2}, orcid=0000-0002-6097-9164]{Nicholas}{Andrews}

\address{
    $^1$ HLTCOE \&
    $^2$ CLSP, Johns Hopkins University, USA \\
    $^3$ Xiaomi Corp., Beijing, China
}

\email{zlin@ieee.org}


\usepackage{comment}
\usepackage{amsmath,hyperref}
\usepackage{multirow, multicol, booktabs}
\usepackage{threeparttable}
\usepackage{enumitem}
\usepackage{xcolor}
\usepackage{appendix,balance}

\begin{document}

\maketitle

\begin{abstract}

Large language models (LLMs), trained on large-scale text, have recently attracted significant attention for their strong performance across many tasks. Motivated by this, we investigate whether a text-trained LLM can help localize fake words in partially fake speech, where only specific words within a speech are edited. We build a speech LLM to perform fake word localization via next token prediction. 
Experiments and analyses on AV-Deepfake1M and PartialEdit indicates that the model frequently leverages editing-style pattern learned from the training data, particularly word-level polarity substitutions for those two databases we discussed, as cues for localizing fake words. Although such particular patterns provide useful information in an in-domain scenario, how to avoid over-reliance on such particular pattern and improve generalization to unseen editing styles remains an open question.

\end{abstract}

\section{Introduction}

Advances in generative artificial intelligence (GenAI) have enabled high-quality synthesis and manipulation of human voices. These technologies can now generate fake speech that is nearly indistinguishable from genuine speech to both human listeners and automated detectors~\cite{alali2025partial}. When only a few words are manipulated while the rest remains authentic, known as ``Partial Fake\footnote{It was known as ``Partial Spoof'' in the authentic scenario~\cite{Zhang2021PartialSpoof}.}''~\cite{Zhang2021PartialSpoof}, detection becomes even harder~\cite{alali2025partial}.
To support research on partially fake speech, several datasets have been released: PartialSpoof~\cite{Zhang2021PartialSpoof}, HAD~\cite{Yi2021halftruth}, LlamaPartialSpoof~\cite{luong2025llamapartialspoof}, PartialEdit~\cite{zhang25g_interspeech}, etc. And some challenges were held: ADD 2022-2023~\cite{yi2022add, yi2023add}, and AV-Deepfake1M~\cite{cai2023avdf1m, cai2025av1m++}. Notably, AV-Deepfake1M~\cite{cai2023avdf1m} is one of the largest available audio-visual partially fake dataset, and the recent PartialEdit dataset~\cite{zhang25g_interspeech} employs advanced speech infilling techniques to edit speech.

In such partial fake scenarios, locating the manipulated part is crucial

For locating the fake regions in the partially fake speech, several approaches have been explored~\cite{he2025manipulated_survey}, which can be grouped into three categories:
(1) frame-level uniform segmentation~\cite{Yi2021halftruth, zhang2022partialspoof, xie2024efficient} which makes per-frame decisions,
(2) frame-level inconsistency~\cite{ wu2024CFPRF, li2025frame} which exploits discrepancies between adjacent frames to make decisions, and
(3) boundary-based localization~\cite{zhong2024BAM, haibin2022partialQA, ge2025gncl}, which focuses on detecting change points (i.e., concatenation parts between bona fide and manipulated segments).
%
Among them, frame-level uniform segmentation is the simplest and most efficient solution. A recent work~\cite{zhang2026wedefense} shows that a self-supervised model with a simple attention-based backend can achieve results comparable to those of more complex solutions for fake localization.

In recent years, LLMs have attracted attention in the nature language processing field~\cite{zhao2023survey}, and they were swiftly adopted in the speech processing once solutions were proposed to support speech inputs through foundational speech representations such as self-supervised learning (SSL) representations~\cite{chen2021wavlm, BaevskiZMA20-w2v2} and neural codecs~\cite{dac, encodec}.
The projection of those speech representations into the embedding space, shared with text, of the LLMs allowed the creation of large speech language model (Speech LLM)~\cite{arora2025landscape_speechllm, peng_survey_2025}.
When generating the partially fake speech, speech editing~\cite{kassmann2024speech} typically targets modify word(s) to change its original meaning, by replacing them with synthetic
words of different semantics meaning.
Then, some specific types of words (like antonym words) could be modified more often.
And some words might be changed to the speaker's unusually used words/extend speakers' speaking style.
Such modification pattern could be serve as cues for localizing fake words.

Two contemporaneous works reformulate defending against partially fake speech as question answering using Speech LLM model.
 Xu \MakeLowercase{\textit{et al.}}~\cite{xu2026holiantispoof} focus more on jointly modeling attack identification, localization, and semantic influence. While Xue \MakeLowercase{\textit{et al.}}~\cite{xue2026unifying} focus on joint detection and localization,  leveraging frame-level probabilities from a conventional frame-level detector. Both demonstrate the potential of speech LLMs for localizing fake words. However, they provide limited analysis of the patterns by which the LLM identifies edited regions within partially fake speech.

\begin{figure*}[!tb]
    \centering
    \includegraphics[width=0.9\linewidth]{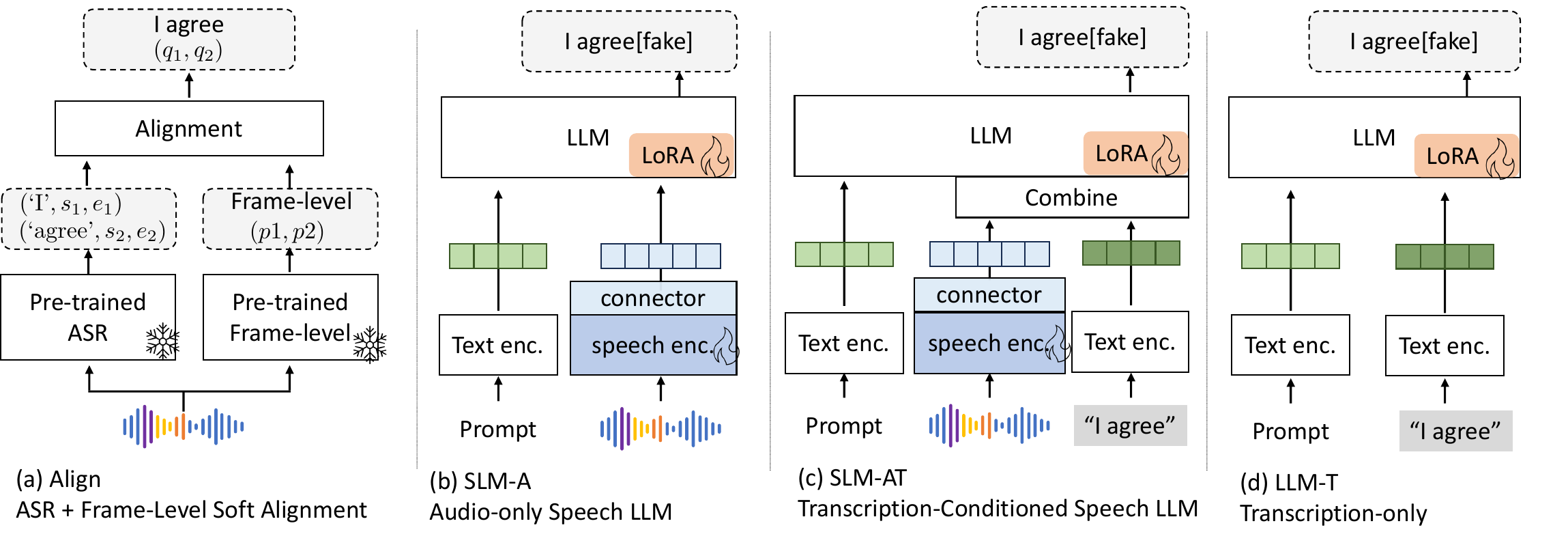}
    \vspace{-3mm}
    \caption{Fake-word localization via (a) Alignment between ASR and frame-level detector, and LLM-based approaches with three modality cases: (b) audio-only, (c) transcription-conditioned audio, and (d) transcription-only.}
    \label{fig:model}
\end{figure*}

Therefore, this study aims to explore not only: \textit{``Can LLMs help localize fake words in partially fake speech?''} but also \textit{``What patterns do they exploit to do so?''}
This study focuses on word-level fake localization to explore this question.
We first build a baseline that aligns ASR-produced words with their timestamps using an SSL-based model for reference. 
Then, we introduce a Speech LLM model to locate fake words via next token prediction.
We consider both settings where transcriptions are unavailable and where they are available.

{Our experiments and analyses on AV-Deepfake1M and PartialEdit suggest that LLMs rely on editing-style patterns they learned from their training data. For example, both databases are edited based on ChatGPT to change meaning. Such editing make fake words usually happen in polarity substitution.
and these patterns differ by modality. With transcription-only as input, models primarily learn \textbf{lexical} cue, like which words are often be edited, at the \textbf{word level}. With audio-only input, speech LLM primarily learns from the \textbf{acoustic} cue, at the \textbf{phoneme level}. 
Although such patterns can aid in-domain fake word localizing to some extent, avoiding over-reliance on dataset-specific patterns and improving generalization to unseen editing styles remain important directions for future work.}

\section{From Alignment to Speech LLM}
As shown in Figure~\ref{fig:model}, this paper mainly focuses on two types of models: our baseline detector based on aligning pretrained ASR and frame-level detector, and speech LLM with different modalities as input.

\subsection{Alignment Baseline}
Figure~\ref{fig:model}(a) shows our baseline detector (\texttt{Align}).
It consists of an ASR and a frame-level real/fake detector. Frame-level detections are aligned to word-level intervals using ASR-derived timestamps.

Given an input $\boldsymbol{x}$, the frame-level detector produces a sequence of posterior scores $\boldsymbol{p}_{1:T} = (p_1, \cdots, p_t, \cdots ,p_T)$, where $p_t$ presents the prediction at the $t$-th frame. Meanwhile, Meanwhile, the pretrained ASR outputs a sequence of words with time stamps $\{w_n, s_{n}, e_{n}\}$, where $w_n$ is the $n$-th word with start time $s_n$ and end $e_n$.
To obtain word-level predictions, we aggregate the frame-level scores within each word’s time span into a per-word score:
\begin{equation}
q_n = \text{Agg}(\{p_t | t \in \mathcal{I}_n\}), \quad \quad \mathcal{I}_n = \{t|t \in [s_n, e_n]\}.
\end{equation}
Where $\text{Agg}(\cdot)$ is the aggregation function, and we use mean aggregation in this study: $q_n = \frac{1}{|\mathcal{I}_n|}\sum_{t\in\mathcal{I}_n}p_t$.

\subsection{Speech LLM for fake word localization}\label{sec:model_speechllm}
As shown in Figure~\ref{fig:model}(b-d), we build a series of speech LLMs based on pre-decoder alignment structure~\cite{peng_survey_2025}, to locate deepfake words through next token prediction. It consists of (i) a speech encoder to produce acoustic embeddings, (ii) a connector module that project the audio embedding to LLM's text feature space, and (iii) LLM.

Given speech and prompt with optional transcription, the LLM will performs next-token prediction to attach ``[fake]'' after any word detected as fake. Similar strategy with next-token prediction for fake word localization has been shown effective in concurrent work by fine-tuning Whisper~\cite{tran2026deepfake}.

Speech LLM for fake word localization serves two purposes: (i) transcribing the speech and (ii) localizing fake words. Accordingly, we consider two scenarios: 
\begin{enumerate}
    \item Transcriptions are unavailable and only audio is provided. Then the LLM need to transcribe and localize fake words.
    \item Transcriptions are available. Then, the task becomes easier, and the LLM only needs to localize fake words, using the speech as additional evidence.
\end{enumerate}
Accordingly, we discussed three input-modality conditions in this study, as illustrated in Figure~\ref{fig:model}(c–d): (b) \texttt{SLM-A} with audio-only as input for scenario 1, (c) Audio+transcription (\texttt{SLM-AT}) for scenario 2 when transcriptions are available. And (d) Transcription-only (\texttt{LLM-T}) is also considered for reference to check how LLMs learn from transcription.

\section{Experimental Setups}
\subsection{Databases}\label{sec:exp_database}
Two databases in different scales are discussed in this paper:

\noindent\textbf{PartialEdit (PE)}~\cite{zhang25g_interspeech} is a recently released partially fake database generated by speech infilling or editing.
PartialEdit is built upon VCTK~\cite{veaux_cstr_2016}, which contains 110 speakers, each with 400 read sentences. These sentences are drawn from newspapers, the rainbow passage, and an elicitation paragraph. 
Consequently, the ground-truth transcripts are available. 
Following AV1M, these transcripts were manipulated using ChatGPT~\cite{achiam2023gpt} to alter the meaning of the sentences, then the speech was edited using a set of synthetic systems. 
Due to licensing constraints, we use only the E1 (VoiceCraft\cite{peng2024voicecraft}) and E2 (SSR\cite{wang2025ssr}) subsets from PartialEdit in this work.





\noindent\textbf{AV-Deepfake1M (AV1M)}\footnote{We didn't use the lastest version AV-Deepfake1M++~\cite{cai2025av1m++} because parts of the validation set metadata are unavailable.}~\cite{cai2023avdf1m} is built upon VoxCeleb2~\cite{chung2018voxceleb2}. 
VoxCeleb2 is a dataset of celebrities' voices collected from YouTube,
originally built for the training speaker verification systems, and primarily comprising spontaneous public interviews and speeches.
When generating partially fake media, the transcriptions were first obtained using Whisper~\cite{pmlr-v202-radford23whisper}, and the target manipulations were generated with ChatGPT~\cite{achiam2023gpt} to revise the meaning in the opposite direction. 
YourTTS~\cite{casanova2022yourtts} and VITS~\cite{kim2021vits} systems are then utilized to generate audio corresponding to the modified words with remaining parts unchanged. 
We extract audio tracks from the AV-Deepfake1M videos. 
AV1M includes four sample types: audio-modified, visual-modified, both-modified, and real. Since we focus on audio analyses and thus use only the audio-modified and real subsets. And since metadata for evaluation set is not yet released, we use only the provided training and validation subsets.


\subsection{Experimental setups}
For the \texttt{Align} model, we require (1) a ASR model to produce transcription with word-level time stamps, and (2) a localization model produces time stamps for fake regions. 
For ASR, we use preained parakeet-tdt-0.6b-v2\footnote{\href{https://huggingface.co/nvidia/parakeet-tdt-0.6b-v2}{https://huggingface.co/nvidia/parakeet-tdt-0.6b-v2}} to get word-level time stamps.
For frame-level fake localization, we adopt SSL-MHFA~\cite{Peng2023AnVerification} given its relatively better performance with lightweight design~\cite{zhang2026wedefense} , with the same configuration as in~\cite{Peng2023AnVerification}.

For the speech LLM model, we largely follow the configuration from ~\cite{thebaud2025enhancing} using their open-sourced code\footnote{ \href{https://github.com/thomasthebaud/speechLLM}{https://github.com/thomasthebaud/speechLLM}}. 
Specifically, we use WavLM-base-plus\footnote{\href{https://huggingface.co/microsoft/wavlm-base-plus}{https://huggingface.co/microsoft/wavlm-base-plus}} as speech encoder, and 
TinyLlama\footnote{\href{https://huggingface.co/TinyLlama/TinyLlama-1.1B-Chat-v1.0}{https://huggingface.co/TinyLlama/TinyLlama-1.1B-Chat-v1.0}} as an LLM component, given its ability with only 1.1B parameters. Such WavLM and Llama combination produce relatively better performance in ASR task~\cite{fortier2025backdoor}. 
Based on our preliminary comparison, two conventional layers are used as a connector. When fine-tuning the speech encoder, the learning rate is set as 2e-06. The learning rate for LLM and the connector is set as 0.0001. 
LoRA is applied with rank=8 and alpha=16.


\begin{table}[!t]
\centering
\begin{threeparttable}
\vspace{6mm}
\caption{Cross-domain WordF1 (\%) using the Align model.}\label{tab:align_cross}
\vspace{-6mm}
\begin{center}
\setlength{\tabcolsep}{10pt}
\begin{tabular}{rccc}
\toprule
\multicolumn{1}{c}{\multirow{2}{*}{Train}} & \multicolumn{3}{c}{Test} \\
                        \cmidrule(r){2-4} 
& PE-dev & PE-eval & AV1M-dev  \\
\midrule
PE-train  & 54.28 & 53.75 & 12.03 \\
AV1M-train & \phantom{0}0.04  & \phantom{0}0.04  & 41.48 \\
\midrule
WER (\%) & \phantom{0}2.20 & \phantom{0}1.94 & 11.99$^\dagger$ \\
\bottomrule
\end{tabular}
     \begin{tablenotes}[flushleft] \setlength{\parindent}{0pt}
       \item \noindent\scriptsize{$\dagger$ Note that WER on AV1M is only used for reference, as its transcription references are generated by whisper but not human annotation~\cite{cai2023avdf1m}. }
     \end{tablenotes}

\end{center}
\end{threeparttable}

\vspace{-3mm}
\end{table}

\begin{table}[t]
\caption{In-domain WER and WordF1 (\%) of LLMs with different input modalities. (a) PE: models trained on PE-train and evaluated on PE-eval. (b) AV1M: models trained on AV1M-train and evaluated on AV1M-dev. }
\label{tab:modality_av1m_pe}
\centering
\scalebox{0.86}{
\begin{tabular}{lcc|cc}
\toprule
\multicolumn{1}{c}{\multirow{2}{*}{Modalities}} & \multicolumn{2}{c|}{(a) PE} & \multicolumn{2}{c}{(b) AV1M}  \\
\cmidrule(lr){2-3} \cmidrule(lr){4-5}
 & WER $\downarrow$ & WordF1 $\uparrow$ & WER $\downarrow$ & WordF1 $\uparrow$ \\
\midrule
\texttt{SLM-A}     & 62.89 & 54.08  & 11.63 & 94.68 \\
\texttt{LLM-T}     &  \phantom{0}0.07 & 83.63  & \phantom{0}0.00 & 84.80 \\
\texttt{SLM-AT}   & \phantom{0}0.02 & 90.79   &  \phantom{0}0.01 & 97.51 \\
\bottomrule
\end{tabular}
}
\end{table}
\subsection{Metrics}
Most prior work on fake word localization emphasizes fine-grained temporal precision, and evaluation metrics are correspondingly designed for fine-grained timestamps as well.
Therefore, we introduce word-level metric, WordF1, used in our study to evaluate word-level fake localization.

Here, we adopt a word-level F1 (WordF1) to measure the model’s ability to detect fake words, treating fake as the positive class. JIWER\footnote{\href{https://jitsi.github.io/jiwer}{https://jitsi.github.io/jiwer}} is used to align words from the reference and the hypothesis based on Levenshtein distance.
In practical attacking scenarios, attackers typically modify a minimal number of words to alter the semantic meaning of speech while maintaining overall naturalness. Thereby the proportion of fake words is significantly lower than that of real words.
Using WordF1 with fake as the positive class provides a more informative evaluation of performance on this minority class and helps avoid artificially high F1 scores obtained by predicting most words as the majority (real) class.

Note that WordF1 is computed over words that are aligned with ASR, assuming they are accurately recognized or that substitutions are aligned to the reference. 
Therefore, the reliability of WordF1 is intrinsically linked to ASR performance. To account for this, we also report the Word Error Rate (WER). 
If WER is too high, word-level localization metrics may cover fewer words due to misalignment between the reference and the hypothesis. 

\section{Results}
\subsection{Fake words localization with the Align model}\label{res:align}
We first construct cross-domain experiments for the \texttt{Align}
in Table~\ref{tab:align_cross}. 
These results serve as a reference to assess whether the speech LLM-based model achieves reasonable performance.

From in-domain experiments in Table~\ref{tab:align_cross}, PE yield lower WER and higher WordF1.
%
From the cross-domain results in Table~\ref{tab:align_cross}, we observed extremely low WordF1 (0.04) when using a model trained on AV1M-train, evaluated on PE-dev and PE-eval. A likely cause is the low fake-word ratio in AV1M, which biases the model toward predicting ``real,'' leading to widespread misses in PE under out-of-domain conditions. This highlights limited generalization in the \texttt{Align} model.

\subsection{Can LLMs help localize fake words in partially fake speech? -- Analyses with different modalities.}

\begin{table}[t]
\caption{Cross-domain WordF1 (\%) using SLM-AT.}\label{tab:cross_AT}
\vspace{-3mm}
\centering
\scalebox{0.9}{
\begin{tabular}{llll}
\toprule
\multicolumn{1}{c}{\multirow{2}{*}{Train}} & \multicolumn{3}{c}{Test} \\
                        \cmidrule(r){2-4} 
 & PE-dev & PE-eval & AV1M-dev \\
\midrule
PE-train  & 91.10 & 90.79   & 32.26    \\
AV1M-train & \phantom{0}0.04  & \phantom{0}0.09    & 97.51    \\
\bottomrule
\end{tabular}
}
\vspace{-3mm}
\end{table}

\begin{table*}[tb]
\centering
\scriptsize
\setlength{\tabcolsep}{5pt}
\renewcommand{\arraystretch}{1.5}
\caption{Top-10 \textbf{fake} lexical words and phoneme cues in the training data and testing data under different modalities of input. Both lists are ordered by frequency. Column IDs are shown within brackets (Column ID).}\label{tab:top10}
\vspace{-3mm}
\begin{tabular}{c p{2.6cm} p{1.7cm} p{2.7cm} p{1.7cm} p{2.7cm} p{1.7cm}}
\toprule
\multicolumn{1}{c}{\multirow{2}{*}{Dataset}} & \multicolumn{2}{c}{Train} 
& \multicolumn{2}{c}{Prediction Pattern on AV1M-dev (\texttt{SLM-A})} 
& \multicolumn{2}{c}{Prediction Pattern on AV1M-dev (\texttt{LLM-T})} \\
\cmidrule(lr){2-3} 
\cmidrule(lr){4-5} 
\cmidrule(lr){6-7}
 & (1) word & (2) phoneme & (3) word & (4) phoneme & (5) word & (6) phoneme \\
\midrule
PE-train
& ignore, a, bad, terrible, ignored, rejected, lost, lose, easy, an
& AH, N, T, D, \underline{IH}, \underline{S}, R, \underline{L}, K, EH, IY, \underline{ER}, P, M, EY 
& \underline{silver}, journey, to, manager, blue, absorption, on, are, you, data
& AH, T, IH, N, S, L, R, ER, D, IY, V, P, K, EH, Z
& terrible, dislike, bad, hate, boring, always, disappointed, do, ordinary, worst
& AH, T, IH, S, L, N, D, R, K, EH, IY, B, P, ER, M
\\
\midrule
AV1M-train
& terrible, do, not, dislike, boring, very, hate, did, definitely, t
& AH, T, IH, N, L, D, S, R, IY, K, EH, M, B, ER, P
& terrible, do, not, dislike, bad, did, hate, boring, very, easy
& AH, T, IH, N, D, L, S, R, IY, K, EH, B, F, M, ER
& terrible, do, dislike, bad, hate, not, did, boring, very, t
& AH, IH, T, D, L, N, S, R, IY, K, EH, B, ER, M, F \\
\bottomrule
\end{tabular}
\vspace{-5mm}
\end{table*}


We explore this question using LLM under different input modalities as we introduced in Section~\ref{sec:model_speechllm}:
with audio-only (\texttt{SLM-A}), transcription-only (\texttt{LLM-T}), or both modalities (\texttt{SLM-AT}). 
In-domain results for PE and AV1M are shown in Table~\ref{tab:modality_av1m_pe} (a) and (b), separately.

We first look at the first row (\texttt{SLM-A}), which accepts audio only, the same input condition as the \texttt{Align} baseline in Section~\ref{res:align}. On PE-eval, \texttt{SLM-A} struggles to produce accurate transcripts (WER = 62.89\%). But under other aligned transcription, it achieves a WordF1=54.08\% comparable to the \texttt{Align} approach, indicating that a speech LLM can localize fake words to some extent.
On AV1M-dev, \texttt{SLM-A} attains a similar WER (11.63\%) to \texttt{Align} but a much higher WordF1 (94.68\%), suggesting that speech LLMs can effectively localize fake words when its transcription quality is adequate. 
This answered our question: Speech LLM can locate fake words in partially fake speech under in-domain scenario, but may limited by transcription quality in some settings.
How to improving audio-only transcription while locating fake words using speech LLM is left for future work.

Therefore, we considered the second scenario introduced in subsection~\ref{sec:model_speechllm}, where transcriptions are available, and the speech LLM only needs to locate fake words.
With transcripts provided, the models (\texttt{LLM-T} and \texttt{SLM-AT}) largely preserves the input text (near-zero WER), indicating minimal alteration or hallucination. Most notably, combining audio with transcripts with \texttt{SLM-AT} consistently improves WordF1 over transcripts only with \texttt{LLM-T}, demonstrating that audio provides complementary cues for locating fake words (WordF1 improved from 84.80\% to 97.51\% on AV1M, and from 83.63\% to 90.79\% on PE-eval). 

We also evaluated the cross-domain using \texttt{SLM-AT} as shown in Table~\ref{tab:cross_AT}. When trained on AV1M and tested on PE, the model yields extremely low WordF1 and tends to label all words as real. A similar pattern appears for the \texttt{Align} baseline in Section~\ref{res:align}, indicating that poor generalization is a general challenge that both methods struggle to address. 

Generalization issues were also reported in other studies when they utilizing Speech LLM based question answering~\cite{xu2026holiantispoof} for fake word localization. This raises the question: why performance of Speech LLM degrades under cross-domain scenarios?

Another notable observation is that even without audio, the model achieves 83.63\% with transcription-only input in the in-domain setting. What patterns do LLMs exploit to localize fake words under different modalities?

\subsection{What patterns do LLMs exploit to localize fake words?}

  
To investigate why performance degrades under cross-domain conditions and how does a LLM localize under different modalities, we conduct a preliminary analysis on the frequency of fake words and the phoneme distributions within those fake words, both in the training data and in the model’s predictions. The top-10 fake words and phonemes are shown in Table~\ref{tab:top10} ordered by their frequency.
Analyses are conducted mainly on the more complicated AV1M-dev. The first row corresponds to models trained on PE-train (cross-domain), and the second row corresponds to models trained on AV1M-train (in-domain). For each training set, we first list frequent words and phonemes in that training data (columns 1–2). We then report statistics from modality-specific models evaluated on the AV1M-dev: \texttt{SLM-A} (Audio-only) predictions in Columns 3–4 and \texttt{LLM-T} (transcription-only) predictions in Columns 5–6.

From Table~\ref{tab:top10} column 1-2, we observe similar editing patterns across both training data: many words are modified toward negative sentiment (e.g., ``bad,'' ``terrible''). This is understandable because, as noted in Section ~\ref{sec:exp_database}, the original speech is largely public-facing and tends to be positive. When prompted to ``inverse its meaning to the opposite direction,''~\cite{cai2023avdf1m}, ChatGPT often substitutes words with their antonyms, frequently replacing positive terms with negative counterparts, thereby altering the overall meaning of the utterance.
These editing patterns are readily learned by \texttt{LLM-T} with transcription-only as input, as reflected in its predictions in the last columns 5-6. 

Similar but distinct patterns are observed from \texttt{SLM-A} prediction as shown in columns 3-4.
When it accepts audio-only as input, the model learns \textbf{phonetic} cues of manipulation.
Among the frequently predicted words, ``silver'' appears with highest frequency, despite not being in the training data's top-10 words. It is likely detected because its phoneme sequnce [S IH L V ER] largely comprises frequent training phonemes (except /V/). This indicates the \texttt{SLM-A} tend to localize fake words by relying on phone-level acoustic evidence.

{
Although these editing patterns can help LLMs localize fake words in-domain to some extent, by capturing how original sentences are modified, over-reliance on them may cause the model to overlook other artifacts and skew it toward locate particular pattern (like antonym or negative words) rather than truly detecting deepfake words. This may explain the degraded performance under cross-domain conditions. 
In real-world settings edits can be more varied and subtle: besides flip the meaning, an attacker might modify named entities.
Over-reliance on particular patterns in the training data can therefore degrade performance in the wild.
Avoiding over-reliance on dataset-specific patterns and improving generalization to unseen editing styles when utilizing LLMs remain important directions for future work.
}

\section{Conclusion}
This study investigated ``Can LLMs help localize fake words in partially fake speech?''
To do so, we built speech LLMs for fake word localization via next token prediction. We considered two scenarios: when transcriptions are unavailable and when they are available.
Our results show that the speech LLM can locate fake words in partially fake speech under in-domain scenario, but degraded under cross-domain scenario.

To understand ``why performance of Speech LLM degrades under cross-domain scenarios?'' and ``how does a LLM helps on localizing fake words?'' we preliminarily analyzed patterns on lexical related words and acoustic related phonemes in the training data and model predictions. 
With transcription-only as input, LLM helps on localizing fake words for in-domain scenario by learning editing patterns from training data, which is often antonym substitutions. And when audio-only as input, speech LLM 
Although such particular patterns may help locate edited words, over-reliance on particular patterns in the training data can therefore degrade performance in the wild.
Avoiding over-reliance on dataset-specific patterns and improving generalization or adaptation to unseen editing styles when utilizing LLMs remain important directions for future work.

\section{Acknowledgment}
The author would thanks Kai Li from Tsinghua University for discussion and valuable suggestions.
This work was supported by the Office of the Director of National Intelligence (ODNI), Intelligence Advanced Research Projects Activity (IARPA), via the ARTS Program under contract D2023-2308110001. 

\section{Generative AI Disclosure}
Generative AI tools were used only for editing and polishing the human-written draft. All AI-assisted text was reviewed by the authors before submission.

\bibliographystyle{IEEEtran}
\bibliography{main}

@article{zhao2023survey,
  title={A survey of large language models},
  author={Zhao, Wayne Xin and Zhou, Kun and Li, Junyi and Tang, Tianyi and Wang, Xiaolei and Hou, Yupeng and Min, Yingqian and Zhang, Beichen and Zhang, Junjie and Dong, Zican and others},
  journal={arXiv preprint arXiv:2303.18223},
  volume={1},
  number={2},
  pages={1--124},
  year={2023}
}

@inproceedings{wu2024CFPRF,
  title={Coarse-to-fine proposal refinement framework for audio temporal forgery detection and localization},
  author={Wu, Junyan and Lu, Wei and Luo, Xiangyang and Yang, Rui and Wang, Qian and Cao, Xiaochun},
  booktitle={Proc. ACM MM},
  pages={7395--7403},
  year={2024}
}

@article{xue2026unifying,
  title={Unifying Speech Editing Detection and Content Localization via Prior-Enhanced Audio LLMs},
  author={Xue, Jun and Chai, Yi and Ren, Yanzhen and He, Jinshen and Tang, Zhiqiang and Yi, Zhuolin and Huang, Yihuan and Xie, Yuankun and Chen, Yujie},
  journal={arXiv preprint arXiv:2601.21463},
  year={2026}
}

@inproceedings{xie2024efficient,
  title={An efficient temporary deepfake location approach based embeddings for partially spoofed audio detection},
  author={Xie, Yuankun and Cheng, Haonan and Wang, Yutian and Ye, Long},
  booktitle={Proc. ICASSP},
  pages={966--970},
  year={2024}
}

@article{li2025frame,
  title={Frame-level Temporal Difference Learning for Partial Deepfake Speech Detection},
  author={Li, Menglu and Zhang, Xiao-Ping and Zhao, Lian},
  journal={IEEE Signal Processing Letters},
  year={2025},
  publisher={IEEE}
}

@inproceedings{ge2025gncl,
  title={{GNCL: A} graph neural network with consistency loss for segment-level spoofed speech detection},
  author={Ge, Zirui and Xu, Xinzhou and Guo, Haiyan and Yang, Zhen and Schuller, Bj{\"o}rn},
  booktitle={Proc. ICASSP},
  pages={1--5},
  year={2025}
}

@inproceedings{zhong2024BAM,
  title     = {Enhancing Partially Spoofed Audio Localization with Boundary-aware Attention Mechanism},
  author    = {Jiafeng Zhong and Bin Li and Jiangyan Yi},
  year      = {2024},
  booktitle = {Proc. Interspeech},
  pages     = {4838--4842},
  doi       = {10.21437/Interspeech.2024-587},
  issn      = {2958-1796},
}

@inproceedings{peng2024voicecraft,
  title={{VoiceCraft: Z}ero-shot speech editing and text-to-speech in the wild},
  author={Peng, Puyuan and Huang, Po-Yao and Li, Shang-Wen and Mohamed, Abdelrahman and Harwath, David},
  booktitle={Proc. ACL},
  pages={12442--12462},
  year={2024}
}

@article{kassmann2024speech,
  title={Speech Editing--a Summary},
  author={K{\"a}ssmann, Tobias and Liu, Yining and Liu, Danni},
  journal={arXiv preprint arXiv:2407.17172},
  year={2024}
}

@article{xu2026holiantispoof,
  title={HoliAntiSpoof: Audio LLM for Holistic Speech Anti-Spoofing},
  author={Xu, Xuenan and Ren, Yiming and Liu, Liwei and Wu, Wen and Li, Baoxiang and Lu, Chaochao and Wang, Shuai and Zhang, Chao},
  journal={arXiv preprint arXiv:2602.04535},
  year={2026}
}

@article{tran2026deepfake,
  title={Deepfake Word Detection by Next-token Prediction using Fine-tuned Whisper},
  author={Tran, Hoan My and Wang, Xin and Ge, Wanying and Liu, Xuechen and Yamagishi, Junichi},
  journal={arXiv preprint arXiv:2602.22658},
  year={2026}
}

@article{achiam2023gpt,
  title={{GPT-4} Technical Report},
  author={Achiam, Josh and Adler, Steven and Agarwal, Sandhini and Ahmad, Lama and Akkaya, Ilge and Aleman, Florencia Leoni and Almeida, Diogo and Altenschmidt, Janko and Altman, Sam and Anadkat, Shyamal and others},
  journal={arXiv preprint arXiv:2303.08774},
  year={2023}
}

@article{fortier2025backdoor,
  title={Backdoor Attacks Against Speech Language Models},
  author={Fortier, Alexandrine and Thebaud, Thomas and Villalba, Jes{\'u}s and Dehak, Najim and Cardinal, Patrick},
  journal={arXiv preprint arXiv:2510.01157},
  year={2025}
}

@inproceedings{casanova2022yourtts,
  title={{YourTTS: T}owards zero-shot multi-speaker tts and zero-shot voice conversion for everyone},
  author={Casanova, Edresson and Weber, Julian and Shulby, Christopher D and Junior, Arnaldo Candido and G{\"o}lge, Eren and Ponti, Moacir A},
  booktitle={Proc. ICML},
  pages={2709--2720},
  year={2022},
}

@inproceedings{kim2021vits,
  title={Conditional variational autoencoder with adversarial learning for end-to-end text-to-speech},
  author={Kim, Jaehyeon and Kong, Jungil and Son, Juhee},
  booktitle={Proc. ICML},
  pages={5530--5540},
  year={2021}
}

@inproceedings{wang2025ssr,
  title={{SSR-speech: T}owards stable, safe and robust zero-shot text-based speech editing and synthesis},
  author={Wang, Helin and Yu, Meng and Hai, Jiarui and Chen, Chen and Hu, Yuchen and Chen, Rilin and Dehak, Najim and Yu, Dong},
  booktitle={Proc. ICASSP},
  pages={1--5},
  year={2025},
  organization={IEEE}
}

@inproceedings{chung2018voxceleb2,
  title     = {{VoxCeleb2: Deep Speaker Recognition}},
  author    = {Joon Son Chung and Arsha Nagrani and Andrew Zisserman},
  year      = {2018},
  booktitle = {{Proc. Interspeech }},
  pages     = {1086--1090},
}

@article{
encodec,
title={High Fidelity Neural Audio Compression},
author={Alexandre D{\'e}fossez and Jade Copet and Gabriel Synnaeve and Yossi Adi},
journal={Transactions on Machine Learning Research},
issn={2835-8856},
year={2023},
}

@inproceedings{dac,
  title={High-fidelity audio compression with improved rvqgan},
  author={Kumar, Rithesh and Seetharaman, Prem and Luebs, Alejandro and Kumar, Ishaan and Kumar, Kundan},
  booktitle={Proc. NeurIPS},
  volume={36},
  pages={27980--27993},
  year={2023}
}

@article{
arora2025landscape_speechllm,
title={On The Landscape of Spoken Language Models: A Comprehensive Survey},
author={Siddhant Arora and Kai-Wei Chang and Chung-Ming Chien and Yifan Peng and Haibin Wu and Yossi Adi and Emmanuel Dupoux and Hung-yi Lee and Karen Livescu and Shinji Watanabe},
journal={Transactions on Machine Learning Research},
issn={2835-8856},
year={2025},
note={}
}

@inproceedings{cai2025av1m++,
  title={{AV-Deepfake1M++: A} large-scale audio-visual deepfake benchmark with real-world perturbations},
  author={Cai, Zhixi and Kuckreja, Kartik and Ghosh, Shreya and Chuchra, Akanksha and Khan, Muhammad Haris and Tariq, Usman and Gedeon, Tom and Dhall, Abhinav},
  booktitle={Proc. ACM MM},
  pages={13686--13691},
  year={2025}
}

@inproceedings{luong2025llamapartialspoof,
  title={LlamaPartialspoof: An llm-driven fake speech dataset simulating disinformation generation},
  author={Luong, Hieu-Thi and Li, Haoyang and Zhang, Lin and Lee, Kong Aik and Chng, Eng Siong},
  booktitle={Proc. ICASSP},
  pages={1--5},
  year={2025},
  organization={IEEE}
}

@inproceedings{thebaud2025enhancing,
  title={Enhancing Dialogue Annotation with Speaker Characteristics Leveraging a Frozen LLM},
  author={Thebaud, Thomas and Lu, Yen-Ju and Wiesner, Matthew and Viechnicki, Peter and Dehak, Najim},
  booktitle={Proc. ASRU},
  year={2025}
}

@inproceedings{zhang25g_interspeech,
  title     = {{PartialEdit: Identifying Partial Deepfakes in the Era of Neural Speech Editing  }},
  author    = {You Zhang and Baotong Tian and Lin Zhang and Zhiyao Duan},
  year      = {2025},
  booktitle = {{Proc. Interspeech}},
  pages     = {5353--5357},
  doi       = {10.21437/Interspeech.2025-942},
  issn      = {2958-1796},
}

@article{zhang2022partialspoof,
   author={Zhang, Lin and Wang, Xin and Cooper, Erica and Evans, Nicholas and Yamagishi, Junichi},
  journal={IEEE/ACM Transactions on Audio, Speech, and Language Processing}, 
  title={The PartialSpoof Database and Countermeasures for the Detection of Short Fake Speech Segments Embedded in an Utterance}, 
  year={2023},
  volume={31},
  number={},
  pages={813-825},
  doi={10.1109/TASLP.2022.3233236}}

@inproceedings{Zhang2021PartialSpoof,
archivePrefix = {arXiv},
arxivId = {2104.02518},
author = {Zhang, Lin and Wang, Xin and Cooper, Erica and Yamagishi, Junichi and Patino, Jose and Evans, Nicholas},
eprint = {2104.02518},
title = {{An Initial Investigation for Detecting Partially Spoofed Audio}},
booktitle = {Proc. Interspeech},
year = {2021},
pages={4264--4268},
doi={10.21437/Interspeech.2021-738}
}

@inproceedings{
cai2023avdf1m,
title={{AV}-Deepfake1M: A Large-Scale {LLM}-Driven Audio-Visual Deepfake Dataset},
author={Zhixi Cai and Shreya Ghosh and Aman Pankaj Adatia and Munawar Hayat and Abhinav Dhall and Tom Gedeon and Kalin Stefanov},
booktitle={Proc. ACM MM},
year={2024},
}

@InProceedings{pmlr-v202-radford23whisper,
  title = 	 {Robust Speech Recognition via Large-Scale Weak Supervision},
  author =       {Radford, Alec and Kim, Jong Wook and Xu, Tao and Brockman, Greg and Mcleavey, Christine and Sutskever, Ilya},
  booktitle = 	 {Proc. ICML},
  pages = 	 {28492--28518},
  year = 	 {2023},
  volume = 	 {202},
  month = 	 {23--29 Jul},
  publisher =    {PMLR},
}

@inproceedings{yi2023add,
  title={{ADD} 2023: the Second Audio Deepfake Detection Challenge},
  author={Yi, Jiangyan and Tao, Jianhua and Fu, Ruibo and Yan, Xinrui and Wang, Chenglong and Wang, Tao and Zhang, Chu Yuan and Zhang, Xiaohui and Zhao, Yan and Ren, Yong and others},
  booktitle={Proc. IJCAI DADA },
  year={2023}
}

@INPROCEEDINGS{yi2022add,
  author={Yi, Jiangyan and Fu, Ruibo and Tao, Jianhua and Nie, Shuai and Ma, Haoxin and Wang, Chenglong and Wang, Tao and Tian, Zhengkun and Bai, Ye and Fan, Cunhang and Liang, Shan and Wang, Shiming and Zhang, Shuai and Yan, Xinrui and Xu, Le and Wen, Zhengqi and Li, Haizhou},
  booktitle={Proc. ICASSP}, 
  title={{ADD} 2022: the first Audio Deep Synthesis Detection Challenge}, 
  year={2022},
  volume={},
  number={},
  pages={9216-9220},
  doi={10.1109/ICASSP43922.2022.9746939}}

@INPROCEEDINGS{haibin2022partialQA,
  author={Wu, Haibin and Kuo, Heng-Cheng and Zheng, Naijun and Hung, Kuo-Hsuan and Lee, Hung-Yi and Tsao, Yu and Wang, Hsin-Min and Meng, Helen},
  booktitle={Proc. ICASSP}, 
  title={Partially Fake Audio Detection by Self-Attention-Based Fake Span Discovery}, 
  year={2022},
  volume={},
  number={},
  pages={9236-9240},
  doi={10.1109/ICASSP43922.2022.9746162}}

@inproceedings{BaevskiZMA20-w2v2,
 author = {Baevski, Alexei and Zhou, Yuhao and Mohamed, Abdelrahman and Auli, Michael},
 booktitle = {Proc. NeurIPS},
 pages = {12449--12460},
 title = {wav2vec 2.0: A Framework for Self-Supervised Learning of Speech Representations},
 volume = {33},
 year = {2020}
}

@article{chen2021wavlm,
  title={Wavlm: Large-scale self-supervised pre-training for full stack speech processing},
  author={Chen, Sanyuan and Wang, Chengyi and Chen, Zhengyang and Wu, Yu and Liu, Shujie and Chen, Zhuo and Li, Jinyu and Kanda, Naoyuki and Yoshioka, Takuya and Xiao, Xiong and others},
  journal={IEEE Journal of Selected Topics in Signal Processing},
  volume={16},
  number={6},
  pages={1505--1518},
  year={2022},
  publisher={IEEE}
}

@inproceedings{Yi2021halftruth,
archivePrefix = {arXiv},
arxivId = {2104.03617v1},
author={Jiangyan Yi and Ye Bai and Jianhua Tao and Haoxin Ma and Zhengkun Tian and Chenglong Wang and Tao Wang and Ruibo Fu},
title={{Half-Truth: A Partially Fake Audio Detection Dataset}},
year=2021,
booktitle={Proc. Interspeech 2021},
pages={1654--1658},
doi={10.21437/Interspeech.2021-930}
}

@article{alali2025partial,
  title={Partial fake speech attacks in the real world using deepfake audio},
  author={Alali, Abdulazeez and Theodorakopoulos, George},
  journal={Journal of Cybersecurity and Privacy},
  volume={5},
  number={1},
  pages={6},
  year={2025},
  publisher={MDPI}
}

@article{zhang2026wedefense,
  title={{WeDefense: A} Toolkit to Defend Against Fake Audio},
  author={Zhang, Lin and Rohdin, Johan and Wang, Xin and Peng, Junyi and Liu, Tianchi and Zhang, You and Luong, Hieu-Thi and Wang, Shuai and Liang, Chengdong and Silnova, Anna and others},
  journal={arXiv preprint arXiv:2601.15240},
  year={2026}
}

@article{he2025manipulated_survey,
  title={Manipulated regions localization for partially deepfake audio: A survey},
  author={He, Jiayi and Yi, Jiangyan and Tao, Jianhua and Zeng, Siding and Gu, Hao},
  journal={arXiv preprint arXiv:2506.14396},
  year={2025}
}

@ARTICLE{peng_survey_2025,
  author={Peng, Jing and Wang, Yucheng and Li, Bohan and Guo, Yiwei and Wang, Hankun and Fang, YanGui and Xi, Yu and Li, Haoyu and Li, Xu and Zhang, Ke and Wang, Shuai and Yu, Kai},
  journal={IEEE Journal of Selected Topics in Signal Processing}, 
  title={A Survey on Speech Large Language Models for Understanding}, 
  year={2026},
  volume={20},
  number={1},
  pages={2-31},
  doi={10.1109/JSTSP.2025.3640535}}

@article{veaux_cstr_2016,
    title = {{CSTR} {VCTK} {Corpus}: {English} {Multi}-speaker {Corpus} for {CSTR} {Voice} {Cloning} {Toolkit}},
    journal = {The Centre for Speech Technology Research (CSTR)},
    author = {Veaux, Christophe and Yamagishi, Junichi and MacDonald, Kirsten},
    year = {2016},
}

@inproceedings{Peng2023AnVerification,
    title = {{An Attention-Based Backend Allowing Efficient Fine-Tuning of Transformer Models for Speaker Verification}},
    year = {2023},
    booktitle = {Proc. SLT},
    author = {Peng, Junyi and Plchot, Oldrich and Stafylakis, Themos and Mosner, Ladislav and Burget, Lukas and Cernocky, Jan},
    pages = {555--562},
}
\balance
\end{document}